\documentclass[aps, prd, reprint, amsmath, amssymb, nofootinbib, twocolumn]{revtex4-1}
\usepackage{hyperref}
\usepackage{cleveref}
\usepackage{OurQuantum}
\usepackage{graphicx}

\newcommand{\JJ}{\gamma}

\begin{document}
\title{Universal Quantum Computation by Scattering in the Fermi-Hubbard Model}
\date{\today}
\author{Ning Bao}
\author{Patrick Hayden}
\author{Grant Salton}
\author{Nathaniel Thomas}
\affiliation{Stanford Institute for Theoretical Physics, Stanford University, Stanford, CA 94305}
\begin{abstract}
The Hubbard model may be the simplest model of particles interacting on a lattice, but simulation of its dynamics remains beyond the reach of current numerical methods. In this article, we show that general quantum computations can be encoded into the physics of wave packets propagating through a planar graph, with scattering interactions governed by the fermionic Hubbard model. Therefore, simulating the model on planar graphs is as hard as simulating quantum computation. We give two different arguments,  demonstrating that the simulation is difficult both for wave packets prepared as excitations of the fermionic vacuum, and for hole wave packets at filling fraction one-half in the limit of strong coupling. In the latter case, which is described by the $t$-$J$ model, there is only reflection and no transmission in the scattering events, as would be the case for classical hard spheres. In that sense, the construction provides a quantum mechanical analog of the Fredkin-Toffoli billiard ball computer.
\end{abstract}
\maketitle

\section{Introduction}
The aim of quantum Hamiltonian complexity theory is to categorize the basic questions of physics by how difficult they are to resolve computationally~\cite{gharibian2014quantum}. 
%
%
Thanks to the Trotter expansion~\cite{trotter1959product} and the phase estimation algorithm~\cite{1995quant.ph.11026K,cleve1998quantum}, simulating the dynamics of a quantum system is often easier than estimating its ground state energy~\cite{kitaev2002classical,aharonov2009power}. Over the past 15 years, increasingly sophisticated methods have been developed for simulating the physics of spin systems and even quantum field theories on quantum computers~\cite{porras2004effective,jordan2012quantum}. In most cases then, the question is not whether a given system can be efficiently simulated but whether that simulation even requires the full power of quantum computation. Free fermion systems, for example, can be efficiently simulated on a classical computer~\cite{2002PhRvA..65c2325T} while free boson systems appear to be hard to simulate classically while falling short of being able to encode arbitrary quantum computations~\cite{aaronson2011computational}.

In recent work, Childs \emph{et al.} demonstrated that a wide class of quantum systems, including the Bose-Hubbard model, is \emph{universal} for quantum computation, in the sense that is is possible to encode arbitrary quantum computations into their dynamics if their interactions are arranged between the vertices of a particular computation-dependent planar graph~\cite{Childs2013}. Since the Bose-Hubbard model can be simulated on a quantum computer and can simulate arbitrary quantum computations, the complexity of simulating the model is therefore \emph{precisely} the power of quantum computation.

In this article, we prove an analogous result for the Fermi-Hubbard model -- a broadly applicable model with relevance to phenomena ranging from the Mott insulator transition to high-temperature superconductivity~\cite{emery1990phase,Essler2005,2007AcPPA.111..409S}. While the techniques of \cite{Childs2013} apply to some systems of anti-commuting scalars, spin is an integral part of the Fermi-Hubbard model and its universality for quantum computation is not resolved by the earlier results. Along the way, we will also establish the universality of the $t$-$J$ model, thus demonstrating that two widely studied and physically relevant fermionic Hamiltonians are universal for quantum computation.

A remarkable variety of strategies have been proposed for realizing quantum computation: the circuit model~\cite{yao1993quantum}, measurement-based quantum computing~\cite{briegel2009measurement}, adiabatic quantum computing~\cite{farhi2000quantum,farhi2001quantum} and topological quantum computation~\cite{freedman2003topological} being the most prominent. Each of these in turn have given rise to a collection of possible realizations, whether in ion traps~\cite{kielpinski2002architecture}, lattices of cold atoms~\cite{monroe2002quantum} or solid state systems~\cite{loss1998quantum,blais2004cavity}. All of those strategies, however, share the common feature that the quantum computation is executed using some form of time-varying control of the system in question. So it is not the intrinsic system dynamics that is universal for quantum computation, but the engineered time-dependent Hamiltonian.

In this work, we are interested in performing arbitrary quantum computations without the use of any time-dependent control. Moreover, we do not permit the Hamiltonian to be tailored to the task at hand: all sites interact in the same way although we allow the sites to be arranged in a computation-specific planar configuration. These restrictions tie the hands of any experimentalist to the point that this approach may be unlikely to yield practical schemes for quantum computation. Our objective, instead, is to assess the inherent computational power of the Fermi-Hubbard model. 
 
Our approach, as in \citet{Childs2013}, is to encode the quantum circuit representing the computation as a planar graph, with the Hamiltonian governing interactions between adjacent vertices. The computation then proceeds by sending a collection of wave packets into the graph, allowing them to scatter, and then observing the transmitted particles. Unlike in \cite{Childs2013}, however, the computation's quantum information is stored not in the location of the wave packet but, rather, in the spin degrees of freedom of the fermions. To implement gates, we adapt a result of DiVincenzo \emph{et al.} on the universality of the controlled Heisenberg interaction~\cite{diVincenzo2000}, emulating controlled interactions using repeated scattering processes.

The article is organized as follows. In ~\cref{setup} we review the Fermi-Hubbard and $t$-$J$ Hamiltonians, the latter an effective description of the Hubbard model at half-filling and strong coupling. In \cref{ops,tps} we analyze one and two particle scattering in the $t$-$J$ model, respectively. The former is used to route wave packets while the latter is used to implement nontrivial unitary transformations. In \cref{univ} we explain how to combine these elements in order to establish universality, bounding the errors incurred from the use of finite-width wave packets in \cref{err}. An alternate approach to universality for the Hubbard model is explained in \cref{hubbardVac} and related extensions are discussed in \cref{extensions}. While we have made an effort to make the article accessible, we rely heavily on the presentation in \cite{Childs2013}, often only explaining the changes that need to be made to their analysis rather than reproducing the discussion from scratch.

\section{The Fermi-Hubbard and $t$-$J$ models}\label{setup}
The Fermi-Hubbard Hamiltonian on a graph $\mathcal{G} = (V,E)$ is given by 
\begin{equation}\label{hamiltonianHubbard}
H= -t \sum_{\sigma \in \{\uparrow,\downarrow\}}\sum_{\{i,j\}\in E} (c_{i\sigma}^\dagger c_{j\sigma}+c_{j\sigma}^\dagger c_{i\sigma})+U \sum_{i\in V} n_{i\uparrow} n_{i\downarrow},
\end{equation}
where $c_i$ are fermionic operators and $n_{i\sigma} \equiv c_{i\sigma}^\dagger c_{i\sigma}$ (no sum) is the number operator~\cite{Essler2005,Childs2013}.  The parameters of this model are $t$ and $U$ (the onsite Coulomb repulsion strength).

The Fermi-Hubbard Hamiltonian near half-filling and in the limit of large positive $U$ can be transformed into the $t$-$J$ Hamiltonian \cite{Essler2005}.  We do this by identifying the Fermi-Hubbard creation operators $c_i^\dagger$ with $t$-$J$ annihilation operators $a_i$, the Fermi-Hubbard half-filling state with the no-particle state of the $t$-$J$ model, and $J=4t^2/U$. Note that this identifies the large $U$ limit with the small $J$ limit.  The new Hamiltonian is then  
\begin{align}\label{hamiltonianA}
H=\sum_{\{i,j\}\in E}P_S\Big[-&t \sum_{\sigma \in \{\uparrow,\downarrow\}} (a_{i\sigma}^\dagger a_{j\sigma} + a_{j\sigma}^\dagger a_{i\sigma})\\
+&  J \left(\vec{S}_i \cdot \vec{S}_j - \frac{n_i n_j}{4}\right) \Big]P_S,\notag
\end{align} 
where $P_S$ is the projector onto the single particle state
\begin{equation}\label{oneSpinProj}
P_S\equiv I-\sum_{i\in V} n_{i\uparrow} n_{i\downarrow}.
\end{equation}
The projector $P_S$ prohibits hopping of two $t$-$J$ particles onto the same site.  The physical interpretation of this condition is that there cannot be a negative number of fermions on a site (interpreting $t$-$J$ excitations as holes in the Fermi-Hubbard model at half-filling).

As emphasized earlier, an important feature of these Hamiltonians is that they are time-independent and have the same form for all lattice sites;  no external input or control is required beyond the design of the graph. The wave packets used for computation are constructed using excitations above the vacuum defined by the operators $c_{i\sigma}$ and $a_{j\sigma}$ for the Hubbard and $t$-$J$ models, respectively. In light of the relationship between the two Hamiltonians, $t$-$J$ wave packets consist of propagating disturbances in the half-filling Hubbard state. Both constructions are universal for quantum computation, but in the main text we will focus on the $t$-$J$ case in order to highlight some interesting divergences with the \citet{Childs2013} analysis. Details of the universality proof based on $c_{i\sigma}$ wave packets can be found in \cref{hubbardVac}. 

Before proceeding, it is important to note that \cite{Childs2013} did actually prove universality for some Hamiltonians built from anticommuting scalars. The most natural way to realize anticommuting scalars using spin-$1/2$ fermions is to polarize all the fermions in the same direction, thereby freezing out the spin degree of freedom. In that case, however, 
the interaction term in \cref{hamiltonianHubbard} will always be zero and the resulting dynamics classically simulatable~\cite{2002PhRvA..65c2325T}. A different strategy, such as the one we present here, is therefore required to achieve universality in the Fermi-Hubbard and $t$-$J$ models.

Some readers may be surprised that the Fermi-Hubbard model is computationally universal in light of the fact that it has a Bethe ansatz solution in one dimension~\cite{lieb1968absence}. There is no serious tension between these facts, however. While the graphs used in the universality proof do consist of long lines interconnected in an intricate pattern, the graphs are irreducibly planar and, therefore, not amenable to the Bethe ansatz solution. (It is a remarkable fact, however, that there do nonetheless exist translationally invariant local Hamiltonians that are universal for quantum computation~\cite{nagaj2008hamiltonian}.)


\section{One-particle scattering}\label{ops}
If we restrict our attention for the moment to a system containing only one particle, then the Hamiltonian contains only hopping terms:
\begin{equation}
H^{(1)} = -t\sum_{\{i,j\}\in E} |i\rangle \langle j| \ox I_\text{spin} = -t A\ox I_\text{spin},
\end{equation}
where $A$ is the adjacency matrix of $\mathcal{G}$, and $H^{(1)}$ denotes the single-particle Hamiltonian.\footnote{Our convention for the sign of the kinetic term is consistent with the condensed matter physics literature but differs from \cite{Childs2013}.}  (The projectors $P_S$ reduce to the identity here.)

This Hamiltonian corresponds to a quantum walk of the wave packet over the graph $\mathcal{G}$. We will focus on graphs that consist of a finite portion connected to one or more semi-infinite lines (the effect of truncating these lines will be discussed in \cref{err}). States that scatter in from one of the semi-infinite lines, through the finite portion, and then out through the semi-infinite lines can be identified as stationary states of the Hamiltonian.
The associated eigenvalue problem was studied in \cite{childs2012levinson} and \cite[Appendix A]{Childs2013}. Each incoming semi-infinite line will henceforth be referred to as a ``rail.''

\begin{figure}[tb!]
\begin{centering}
\includegraphics[width=0.8\columnwidth]{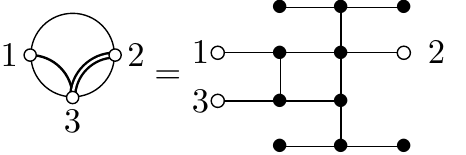}
\par\end{centering}
\caption[A momentum switch]{\label{momentumSwitchGraph}\textbf{A simple momentum switch subgraph.} This subgraph implements a unitary operation corresponding to a momentum switch on a single wave packet.  Additional external semi-infinite graphs are adjoined at the white nodes and carry input and output wave packets.  Input wave packets are directed onto one of two output rails depending on their momenta.  This graph gives rise to a scattering matrix with perfect reflection or transmission, depending on the momentum of the wave packet. This particular momentum switch discerns perfectly between momenta $|k_1|=\pi/4$ and $|k_2|=\pi/2$.}
\end{figure}

A one-particle scattering problem of particular importance in the present context is scattering through a graph known as a ``momentum switch.''  Momentum switches are subgraphs that are engineered to shunt single particle wave packets based on their momenta \cite{Childs2013,childs2014momentum}.  For particular incoming momenta, these subgraphs have perfect transmission or reflection between any two ``input / output" nodes.  There are infinitely many such momentum switch designs, but we will only need one for our construction.  In particular, we choose the subgraph shown in \cref{momentumSwitchGraph}, which was introduced and analyzed in \cite{Childs2013}.  For this graph, wave packets with momentum $\pi/4$ have perfect transmission between nodes 1 and 3, and wave packets with momentum $\pi/2$ have perfect transmission between nodes 2 and 3, where the minus sign arises as a choice of convention.  This momentum switch was designed for spinless particles but functions identically in systems with spin, since the extra spin degree of freedom has no effect on one-particle scattering.

\section{Two-particle scattering}\label{tps}
\label{twoParticleScattering}
Using the momentum switches described in the previous section, we can route wave packets of different momenta toward one another. The interaction term in the Hamiltonian will then induce nontrivial scattering of the wave packets. Our strategy for building unitary gates will be to selectively route pairs of wave packets into an interaction region, taking care to ensure that there are never more than two particles present in the region at a given time.

Throughout the paper we work in one of two bases: the \textbf{uncoupled basis}
\begin{equation}
\{\ket{\uparrow\uparrow},\ket{\uparrow\downarrow},\ket{\downarrow\uparrow},\ket{\downarrow\downarrow}\},
\end{equation}
and the \textbf{coupled basis}
\begin{align}
&\left\{\ket{\uparrow\uparrow},\frac{\ket{\uparrow\downarrow}+\ket{\downarrow\uparrow}}{\sqrt{2}},\frac{\ket{\uparrow\downarrow}-\ket{\downarrow\uparrow}}{\sqrt{2}},\ket{\downarrow\downarrow}\right\}\\
\equiv&\left\{\ket{T_+},\ket{T_0},\ket{S},\ket{T_-}\right\}\notag
\end{align}
In the limit of infinitely long wave packets, scattering of two particles in this interaction region induces a unitary $U$ on the two-spin subspace that will, in general, entangle the two spins.

To see how this unitary arises, we first note that we can write the $t$-$J$ Hamiltonian on a line (restricted to the two-particle subspace) as 
\begin{align}
H^{(2)}=&P_S\Big[-t\left(H^{(1)}_x \ox I_{y}+I_{x}\ox H^{(1)}_y \right) \ox I_\text{spin} \\
  &+ \sum_{x,y\in \ZZ}\proj{x,y} \ox \hat{V}\left(\abs{x-y}\right)\Big]P_S,\notag
\end{align}
where
\begin{equation} \label{kinetic}
H^{(1)}_x \equiv \sum_{x\in \ZZ}\left(\ketbra{x}{x+1}+\ketbra{x+1}{x}\right).
\end{equation}
The potential for the $t$-$J$ model is given by 
\begin{align*}\hat{V}\left(\abs{r}\right) &= J\delta_{\abs{r},1} \left[ (\vec{S}_1 \ox I_2) \cdot (I_1 \ox \vec{S}_2) - \frac{1}{4} I_1 \ox I_2 \right] \\ &=  -\frac{J\delta_{\abs{r},1}}{2} \proj{S}
\end{align*} (recall that $\ket{S}$ denotes the spin singlet state).  
Rescaling the Hamiltonian such that $t=1$, we can rewrite $H$ as 
\begin{align}\label{H2sr}
H^{(2)}=&P_S\Big[-\left(H^{(1)}_s \ox H^{(1)}_r \ox I_\text{spin}\right) \\
  &-J I_s\ox \sum_{r\in \ZZ}\delta_{|r|,1}\proj{r} \ox \proj{S}\Big]P_S,\notag
\end{align}
where we have introduced new variables\footnote{Note that the map $(x,y)\mapsto(r,s)$ is an injection from $\ZZ\times\ZZ$ to a proper subset of $\ZZ\times\ZZ$; we require both $s+r$ and $s-r$ to be even, since our valid lattice sites are those where $x$ and $y$ are integers.  Thus, our original Hilbert space $\mathcal{H}_x\ox\mathcal{H}_y$ is a subspace of $\mathcal{H}_r\ox\mathcal{H}_s$.  When transforming the Hamiltonian into the form of \cref{H2sr}, we arrived at the tensor product structure of $H^{(1)}_s\ox H^{(1)}_r$ by adding irrelevant terms to the Hamiltonian which correspond to non-integer $x$ and $y$.  These terms are always zero when acting on valid (integer $x$ and $y$) input states, and can be added freely.} $r\equiv x-y$ and $s\equiv x+y$.

In terms of the new variables, 
\begin{equation}
P_S=I-\proj{r=0}\ox(\proj{S}+\proj{T_0}),
\end{equation}
which annihilates states with two quasi-particles on the same site.  Note that the action of this projector is equivalent to the projector $(I - {\proj{0}}_r) \otimes I_{\text{spin}}$ on the subspace corresponding to identical fermions.  The potential term in $H^{(2)}$ is unchanged by the projections, while the kinetic terms vanish for transitions to and from $r=0$.  

Now suppose we have two wave packets with momenta $k_1$ and $k_2$ traveling toward one another in some region of the graph.  We define $p_1\equiv -(k_1+k_2)$ and $p_2\equiv (k_2-k_1)/2$.  A generic initial two-particle wavefunction can be written as 
\begin{equation}
\ket\Psi=\sum_{i \in \{\pm,0\}}\ket{\psi_{\text{antisym},i}} \ox \ket{T_i} + \ket{\psi_{\text{sym}}}\ox\ket{S}.
\end{equation}
Antisymmetry of fermions requires that $\psi_{\text{antisym},i}(s,-r) = - \psi_{\text{antisym},i}(s,r)$ and $\psi_\text{sym}(s,-r) = \psi_\text{sym}(s,r)$.  We can view this as a one-particle scattering problem, where we write the wavefunction as
\begin{equation}
\psi(s,r)\equiv e^{-i p_1 s/2} \phi(r)
\end{equation}
and the Hamiltonian as
\begin{equation}
H^{(2)}= -2 \cos \left(\frac{p_1}{2}\right) H^{(1)}_{r\neq 0} \ox I_\text{spin} -J\sum_{r\in \ZZ}\delta_{|r|,1}\proj{r} \ox \proj{S},
\end{equation}
where $H^{(1)}_{r\neq 0}$ is the usual kinetic term without transitions to and from $r=0$.

Two helpful simplifications have been made:  first, the two-particle problem has been reduced to the problem of a single particle scattering off of a potential.  Second, we can now solve the singlet and triplet sectors separately.   Scattering for the triplet sector is trivial because the interaction term is zero.

For the $\ket{S}$ sector, the scattering problem reduces to solving for the eigenvector in an equation of the form $H^{(2)}\phi(r) + \phi_0 = E \phi(r)$ on the two nontrivial sites $r~\in~\{-1,1\}$; outside of these sites, the solution is simply an infinite plane wave in this approximation.  The wavefunction is zero at $r=0$ because we cannot have two $t$-$J$ quasi-particles occupying the same site.  We add the $\phi_0$ term to capture the kinetic contribution of the $r=\pm2$ sites to the wavefunction of the $r=\pm1$ sites.  This yields the following equation:
\begin{align}
&J\left(\begin{array}{c} \notag
e^{i p_2} + R e^{-i p_2}\\
T e^{-i p_2} \end{array} \right)
+2\cos \left(\frac{p_1}{2}\right)\left(\begin{array}{c} 
e^{2 i p_2} + R e^{- 2 i p_2}\\
T e^{- 2 i p_2} \end{array} \right) \\ 
&= 4\cos\left(\frac{p_1}{2}\right)\cos(p_2) 
\left(\begin{array}{c} 
e^{i p_2} + R e^{-i p_2}\\
T e^{-i p_2} \end{array} \right)
\end{align}
where $T$ and $R$ (functions of $p_1$, $p_2$, and $J$) are the coefficients of transmission and reflection.  This scattering problem has the solution
\begin{align}
T&=0\\
R&=-e^{2 i p_2}\left(\frac{J-2\cos(p_1/2)e^{-i p_2}}{J- 2\cos(p_1/2)e^{i p_2}}\right). \notag
\end{align}

\begin{figure}[tb!]
\begin{centering}
\includegraphics[width=0.95\columnwidth]{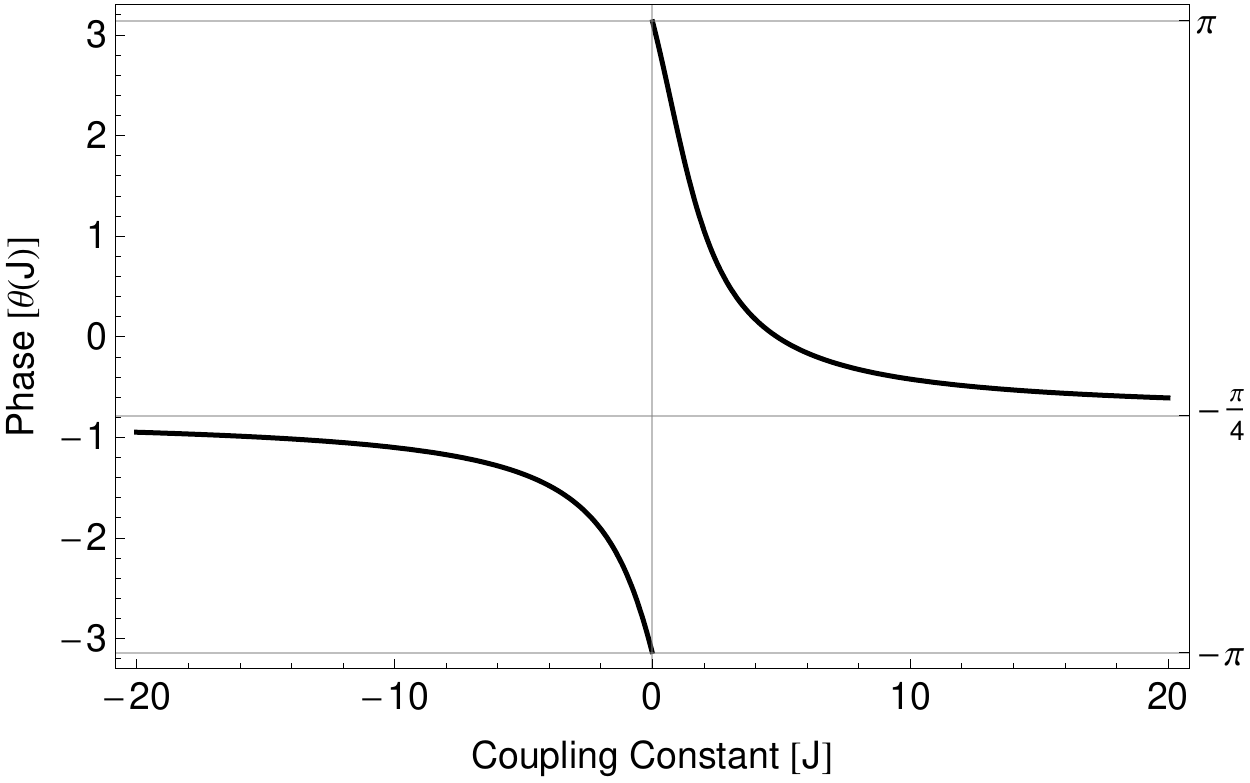}
\par\end{centering}
\caption[$\theta(J)$]{\label{thetaJ}\textbf{Scattering-induced unitary $g$ of \cref{gate-g}.} The phase $\theta$ as a function of the interaction strength $J$ in the $t$-$J$ model.}
\end{figure}

Thus, in the $t$-$J$ model, wave packets collide elastically and this scattering induces a unitary in the two-spin space that is diagonal in the \emph{coupled} basis:
\begin{equation} \label{gate-g}
g=\left(\begin{array}{cccc} 
1 & 0 & 0 & 0  \\
0 & 1 & 0 & 0  \\
0 & 0 & e^{i\theta(p_1,p_2,J)} & 0 \\
0 & 0 & 0 & 1 \end{array} \right),
\end{equation}
where $\theta(p_1,p_2,J) \equiv \text{Arg}(R)$ is plotted in \cref{thetaJ}.

\begin{figure}[!tb]
\begin{centering}
\frame{\includegraphics[width=.2924\columnwidth]{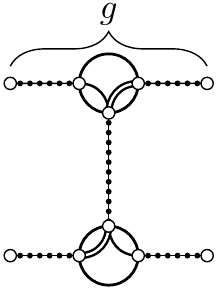}}\quad
\frame{\includegraphics[width=.48\columnwidth]{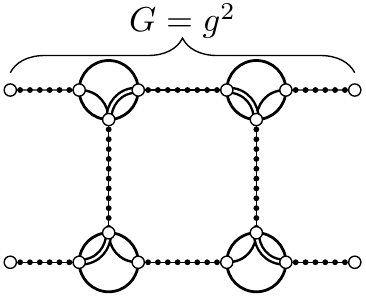}}\end{centering}
\caption[Collision gate unitaries]{\label{collisionGates}Two-particle scatterings which implement the unitary gates $g$ \textbf{(left panel)} and $G=g^2$ \textbf{(right panel)}.  In the left panel, a single scattering collision implements $g$ on the spins of the two rails, but the output rails have different momenta than the incoming wave packets for each rail. Two copies of the gate $g$ are needed to ensure the incoming and outgoing momenta on each rail are the same, as shown in the right panel.  Adding extra lattice sites (not shown) in the appropriate locations ensures that  wave packets exit $G$ simultaneously whenever they entered simultaneously.}
\end{figure}

In order to route two particles toward one another in some interaction region, we make use of the momentum switches described in \cref{ops}.  The unitary $g$ is shown in the left panel of \cref{collisionGates}.
However, a single round of momentum switches results in the wave packets exiting the interaction region on opposite rails.  In order to ensure that wave packets on neighboring rails can always be scattered off of each other, it is convenient to  define our basic unitary gate to be $G=g^2$, as shown in the right panel of \cref{collisionGates}, such that the incoming and outgoing wave packets on each rail have the same momenta.  Extra lattice sites are included as necessary to ensure the two wave packets enter momentum switches and exit the subgraph at the same time.

\section{Universality}\label{univ}
We have seen how single particle scattering through a momentum switch can be used to shunt wave packets  depending on their momenta, and how two-particle scattering can be used to introduce a unitary transformation on the spin subspace of the two colliding wave packets.  We now aim to build a universal quantum gate set using only these two processes.

Following \Citet{diVincenzo2000}, each logical qubit will be encoded in the $S=1/2$, $S_z = +1/2$ subspace of three physical rails.  Our approach differs from that of Childs \emph{et al.}, who employed a dual-rail encoding but generated single qubit unitaries using graph configurations adjoined to single rails.  In our scheme, such single-rail ``gadgets'' only introduce global phases; effecting nontrivial logical gates on single qubits in the triple-rail encoding, requires two-particle scattering between the wave packets of at least two rails.

\subsection{State Preparation}
We will use the same logical (qubit) basis as \cite{diVincenzo2000}:
\begin{align}
\ket{0_L}=&\ket{S}\ket{\uparrow} \\ 
\ket{1_L}=&\sqrt{\frac{2}{3}}\ket{\uparrow\uparrow\downarrow}-\frac{1}{\sqrt{3}}\ket{T_0}\ket{\uparrow}. \notag
\end{align}
These states form a basis for the subspace of total spin quantum number $S=1/2$ and total $S_z=+1/2$.

In order to ensure we can always implement momentum switches between neighboring rails, we initialize alternating rails with square wave packets of momenta $\pi/4$ and $\pi/2$, such that odd numbered rails carry wave packets with low momentum while even numbered rails carry high momentum wave packets.

For the purposes of computation, it is sufficient to prepare the $\ket{0_L}$ state. One way to do so would be to prepare three unentangled wave packets $\ket{\uparrow \downarrow \uparrow}$ and then use repeated $t$-$J$ scattering of the first two rails to approximate the gate $g$ with $\theta = \pi/2$, resulting in the state $(\ket{\uparrow \downarrow} - i \ket{\downarrow \uparrow} )\ket{\uparrow} / \sqrt{2}$. Sending the first rail through a region with an appropriate localized but time-independent magnetic field could then eliminate the unwanted phase, producing the initial state $\ket{S}\ket{\uparrow}$.\footnote{Technically, the presence of the localized magnetic field violates  our prescription of having the same Hamiltonian act at every site, but that is inevitable at the preparation stage. The rest of the computation proceeds using only the unmodified Hamiltonian.}

\subsection{One and two qubit unitaries}
\begin{figure}[!tb]
\begin{centering}
\includegraphics[width=0.95\columnwidth]{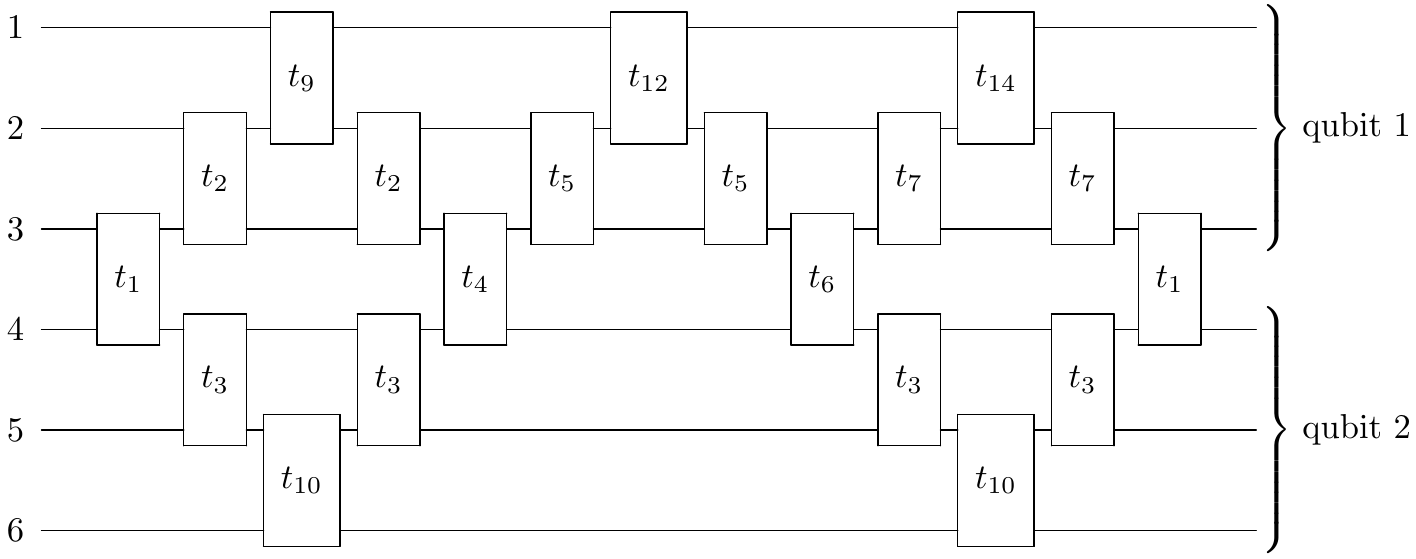}
\par\end{centering}
\caption[One collision gate]{\label{CNOTexample}\textbf{CNOT.}  The encoded CNOT gate from \cite{diVincenzo2000}, as a pattern of Heisenberg interaction times. Adapted for use in the $t$-$J$ model, each of the constituent unitary gates should be interpreted as many repeated scattering gates $G$, as depicted in \cref{collisionGates}, such that the total effect reproduces the unitary of \cref{baconSS} with $t=t_j$. An analogous and simpler construction acting only on the three rails of a single logical qubit suffice to implement general single qubit operations.}
\end{figure}

\Citet{diVincenzo2000} showed that the exchange interaction is universal when it can be dynamically controlled.  In particular, the authors established universality of a unitary spin-spin interaction of the form 
\begin{align}\label{baconSS}
\tilde{U}\equiv e^{i \JJ t \vec{S}_1 \cdot \vec{S}_2} &= \exp \left[i\frac{\JJ t}{4} \right]\exp\left[-i \JJ t\proj{S}\right],
\end{align}
where $t$ is a controllable time parameter.   They used a computer search to determine interaction durations sufficient to produce a universal gate set, finding that a universal set of gates can be produced by systematically interacting qubits for one of fourteen numerically determined times, as needed.  In order to produce a CNOT with matrix elements accurate to at least $6 \times 10^{-5}$, for example, they found that it was sufficient to control $t$ to a precision of $2 \times 10^{-6}$ in the configuration illustrated in \cref{CNOTexample}.

In our case, the interaction timescale is determined by the wave packet width and momentum. In other words, unitary transformations of the spin subspace are implemented in discrete steps corresponding to individual collisions rather than through continuous time evolution.  In order to reproduce a given unitary, two wave packets need to be repeatedly scattered off one another until the desired unitary is achieved.

We can incorporate the approach of \cite{diVincenzo2000} in our construction as follows. Suppose the product $\JJ t$ is specified and we wish to reproduce the corresponding unitary $\tilde{U}$ using only our gate $G = g^2$.  That is, we wish to find a $k$ such that $\| G^k - \tilde{U} \| \leq \epsilon$. In light of \cref{gate-g} and \cref{baconSS},
\begin{equation}
\left\| G^k - \tilde{U}  \right\| 
= | \exp [ i ( 2k\theta(J) ]  - \exp( i \gamma t ) ] |
\leq | 2k \theta(J) - \gamma t |, \nonumber
\end{equation}
so it suffices to approximate $\gamma t$ to within $\epsilon$.
For some values of $\theta(J)$ this will not even be possible, but generically it will be. 

One excellent choice is to tune $J$ such that $\theta(J)/\pi$ is equal to the inverse of the golden ratio, $\phi^{-1} = (\sqrt{5}-1)/2 = 0.618 \ldots$ The sequence of points $1, e^{i 2 \theta}, e^{i 4 \theta}, e^{i 6 \theta}, \ldots, e^{i k 2 \theta}$ will subdivide the unit circle into $k$ intervals. For each $k$, the next point $e^{i (k+1) 2 \theta}$ will  always subdivide the largest of those intervals, with the ratio of the lengths of the two new intervals itself given by the golden ratio~\cite{swier1958circle,knuth1998art}. This procedure therefore distributes successive points very effectively around the circle. In particular, for any $\gamma t$ and $\epsilon$, there will be a $k = O(1/\epsilon)$ such that $\| G^k - \tilde{U} \| \leq \epsilon$.

More generally, the choice of $J$ can be guided by the properties of the continued fraction expansion of $\alpha = \theta(J)/\pi$.  Let $p/q$ be a fraction expressed in lowest terms ($\operatorname{gcd}(p,q)  =  1$). Then the points $e^{i \pi k p / q}$, for $1 \leq k \leq q$, will be uniformly distributed around the unit circle and, therefore, capable of expressing any angle $\gamma t$ to precision $2 \pi / q$. Now suppose that $p/q$ appears as a convergent in the continued fraction expansion of $\alpha$. In that case, $| \alpha - p/q | < 1/q^2$~\cite{miller2006invitation} so
\begin{equation}
\left| k 2 \theta(J) - \pi k p / q \right| 
= k \pi \left| 2 \theta(J)/\pi - p / q \right| 
< \frac{k\pi}{q^2} 
\leq \frac{\pi}{q}. \nonumber
\end{equation}
It follows that ${k 2 \theta(J)}$ for $1 \leq k \leq q$ will be able to approximate any angle to within $2 \pi / q + \pi / q = 3\pi/q$. We should therefore aim to choose $q = O(1/\epsilon)$. Achieving precision $\epsilon$ with $O(1/\epsilon)$ points is, of course, essentially optimal. 

In summary, for any given $J$ and precision $\epsilon$, the question is whether a convergent with  $q = O(1/\epsilon)$  appears in the continued fraction expansion of $\theta(J)/\pi$. Whenever it does, our method can approximate the target $\tilde{U}$ to the desired precision using a number of scattering events scaling inversely with the precision. The circumstances in which this won't be the case are of two types. If $\theta(J)/\pi$ is a rational number with denominator $q$ smaller than $1/\epsilon$, then $G^k$ will be periodic with period $q$ and unable to achieve the desired precision. More interestingly, if the denominator $q$ in the continued fraction expansion grows too rapidly, then the available convergents may be too large and the procedure outlined here will achieve even better precision than necessary but at the cost of additional scattering events. This wouldn't affect universality, but it would impose an undesirable additional constant overhead on the quantum computation. The suitability of a particular $\theta(J)$ can be checked very quickly, however, since one has $q \geq 2^{(r-1)/2}$  for the $r$th convergent~\cite{miller2006invitation}. It is therefore sufficient to calculate the first $\lceil 2 \log_2(1/\epsilon) + 1 \rceil$ convergents and verify if any satisfy $q = O(1/\epsilon)$.

\subsection{Measurement}

A simple and sufficient procedure for measuring the logical state of the triple rail qubit is to measure the third spin along the $z$ direction. If the state is $\ket{0_L}$ then the outcome will always be $\ket{\uparrow}$, but if the state is $\ket{1_L}$ then the outcome will be $\ket{\downarrow}$ with probability $2/3$. Repeating the entire computation and using majority voting can then shrink the error probability exponentially as a function of the number of repetitions.

\section{Error analysis for finite-length wave packets}\label{err}
Our analysis thus far has been exact for wave packets of infinite length.  We now relax this condition to wave packets with finite support in order to show that our scheme is universal with finite wave packets of length polynomial in the number of gates and qubits.  To do so, we can extend Theorems 1 and 2 in \citet{Childs2013}, which respectively control the errors incurred by using finite width wave packets in one- and two- particle scattering of spinless particles, to the $t$-$J$ model.

Theorem 1, in fact, requires no modification because the single particle sector of the $t$-$J$ Hamiltonian contains only the kinetic term, with spin playing no part. In the case of Theorem 2, inspection of the original argument reveals that the reasoning applies \emph{mutatis mutandis} in our case because our Hamiltonian $H^{(2)}$ is only nontrivial on $\ket{S}$, so that the spin degree of freedom is effectively one-dimensional. We conclude that
\begin{align*}
&\Matnorm{e^{-iH^{(2)}t} \ket{\psi(0)} - \ket{\alpha(t)}} \\
&=\Matnorm{e^{i \theta(p_1,p_2,J)} \ket{\psi_\text{sym}(0)}  - \ket{\alpha_\text{sym}(t)}} \\
&= \mathcal{O}(L^{-1/4}),
\end{align*}
where $\psi_\text{sym}$ and $\alpha_\text{sym}$ are, respectively, the symmetric parts of the actual and reference spatial wavefunctions, as described in \cite{Childs2013} and $L$ is the width of the wavefunction.  (Note that there are also $\ket{T_+}$, $\ket{T_0}$, and $\ket{T_-}$ branches of the wavefunctions, but these are equal and will cancel in the norm difference above.)  Thus, both Theorems 1 and 2 in \cite{Childs2013} hold for the $t$-$J$ model, as does the truncation lemma (their Lemma 2).  In particular, the operator norm of the Hamiltonian appears in the truncation lemma, and is $\Matnorm{H} = \mathcal{O}(n^2)$ where $n$ is the number of logical qubits in our case. 

Similarly to \cite{Childs2013}, we find a factor of $\mathcal{O}(n\Matnorm{H}L^{-1/4})$ for the error from each application of $G$: the $L^{-1/4}$ comes from \cite{Childs2013} Theorems 1 and 2;  the $\Matnorm{H}$ is added from application of the truncation lemma; and the $n$ comes from the fact that there may be $\mathcal{O}(n)$ swaps necessary, using the triangle inequality each time.  We have one block for each of the $G$ gates we apply.  Once again utilizing the triangle inequality, we find that the error scales as $\mathcal{O}(mn\Matnorm{H}L^{-1/4})=\mathcal{O}(mn^3 L^{-1/4})$, as in \citet[Equation 19]{Childs2013}, with $m$ the number of logical gates in the computation.  Therefore, we find the same error bounds as Childs \emph{et al.}:  $L = \mathcal{O}(n^{12} m^{4})$, $\mathcal{O}(n^{13} m^{5})$ vertices, and $\mathcal{O}(n^{12} m^{5})$ total evolution time.  We suspect that these bounds are overly conservative and that further analysis could significantly reduce the degree of these polynomials.

\section{Universality in the Dilute Limit} \label{hubbardVac}

Similar universality arguments to those given for the $t$-$J$ model, which governs the Hubbard model at half-filling in the limit of large positive $U$, apply to the Hubbard model itself. In that case, the wave packets are excitations prepared above the fermionic vacuum, as defined by the Hubbard model annihilation operators $c_{i\sigma}$. We will refer to this as the dilute limit, although ``perfectly dilute'' may be a more appropriate term since the only excitations present are those we introduce intentionally.

In the two-particle sector, the Hubbard Hamiltonian \cref{hamiltonianHubbard} can be written as
%
\begin{align*}
H^{(2)} = &-\Big(H^{(1)}_x \ox I_y + I_x \ox H^{(1)}_y\Big) \ox I_{\text{spin}} \\
&+ U \sum_i \proj{ii} \ox \Big( \proj{T_0} + \proj{S} \Big)
\end{align*}
where $H^{(1)}$ is the kinetic term \cref{kinetic}, and $t=1$.
As before, rewriting in terms of $r\equiv x-y$ and $s\equiv x+y$ and using the ansatz where the $s$ and $r$ parts of our wavefunction are separable with the $s$ part equal to $e^{-ip_1 s/2}$, we can simplify to 
\begin{align*}
H^{(2)} = &-2 \cos \left( \frac{p_1}{2} \right) H^{(1)}_r \ox I_{\text{spin}} \\
&+ U \proj{r=0} \ox \Big( \proj{T_0} + \proj{S} \Big).
\end{align*}
Since $\proj{T_0}$ is symmetric, to antisymmetrize the wavefunction the corresponding spatial part of $T_0$ must be antisymmetric.  But note that at $r=0$, such a wavefunction must vanish.  Therefore we can ignore the $T_0$ part of the Hamiltonian, which then simplifies to \[H^{(2)} = -2 \cos \left( \frac{p_1}{2} \right) H^{(1)}_r \ox I_{\text{spin}} + U \proj{0} \ox \proj{S}.\] \\

\begin{figure}[tb!]
\begin{centering}
\includegraphics[width=0.95\columnwidth]{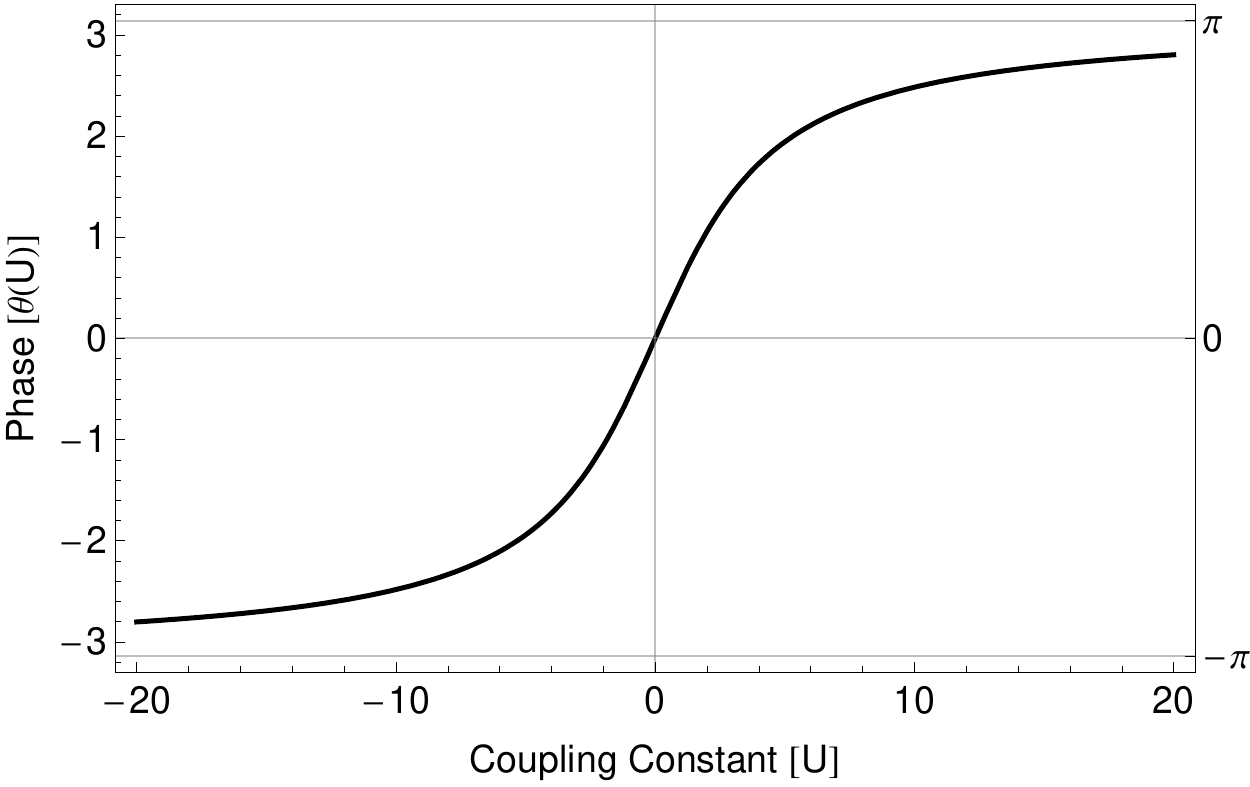}
\par\end{centering}
\caption[$\theta(U)$]{\label{thetaU}\textbf{Scattering-induced unitary $g$ in the Fermi-Hubbard model.} The phase $\theta$ as a function of the interaction strength $U$ in the Fermi-Hubbard model.}
\end{figure}

If we restrict to the $\ket{S}$ sector, the scattering problem becomes identical to the one solved in Appendix B of \citet{Childs2013}  for the Bose-Hubbard model, since the spatial part of the fermionic wavefunction must be symmetric in the $\ket{S}$ sector.  The other sectors have no interaction and therefore trivial scattering.  The scattering produces a phase in the $\ket{S}$ sector defined by the following equation relating the transmission and reflection coefficients: \[T+R = 1 - \frac{2U}{U+4i \cos(p_1 / 2) \sin(p_2)}.\]  As a result, two-particle scattering induces an entangling unitary operator of the same form as \cref{gate-g}, with the phase a function of $U$ rather than $J$, as shown in \cref{thetaU}. As for the $t$-$J$ model, repeated use of such an operator can be used to simulate controlled Heisenberg interactions and thereby build universal quantum computation.
\section{Further Extensions} \label{extensions}

Minor modifications of the arguments presented here can be used to show that an even wider class of fermionic Hamiltonians is universal for quantum computation including the non-isotropic $XXZ$ version of the $t$-$J$ Hamiltonian:
\begin{align}\label{hamiltonianXXZ}
H=\sum_{\{i,j\}\in E}P_S\Big[-&t \sum_{\sigma \in \{\uparrow,\downarrow\}} (a_{i\sigma}^\dagger a_{j\sigma} + a_{j\sigma}^\dagger a_{i\sigma})\\
+&  J_x \left(S_{xi} S_{xj} + S_{yi} S_{yj} \right) + J_z S_{zi} S_{zj} \Big] P_S, \notag
\end{align} 
where $P_S$ is defined in \cref{oneSpinProj}.  The transformation induced by two-particle scattering, analogous to the $g$ of \cref{gate-g} has the form
\begin{align}
\tilde{g} = \left(\begin{array}{cccc} 
1 & 0 & 0 & 0 \\
0 & e^{i \theta_1} & 0 & 0 \\
0 & 0 & e^{i \theta_2} & 0 \\
0 & 0 & 0 & 1   \end{array} \right), \label{new-g}
\end{align}
in the coupled basis. By taking powers of $\tilde{G} = \tilde{g}^2$, the second term on the diagonal can either be made to be exactly 1 if $\theta_1/\pi$ is rational, or to approximate it otherwise. In the first case, the universality of the specific $(\theta_1,\theta_2)$ pair reduces to analyzing universality of a gate $G$ with the corresponding multiple of $\theta_2$. In the second case it is necessary to contend with approximations, but one could again use continued fractions to estimate the size of any errors and the number of scattering events required.

\section{Discussion}

The fermionic Hubbard model, despite its simplicity, captures many essential features of the physics of electrons in solids. In this article, we have shown that that rich variety of behavior extends to universal quantum computation: simulating the Hubbard model on an arbitrary graph, both just below half-filling and in the dilute limit, is as hard as simulating arbitrary quantum computers. The graph itself encodes the computation to be performed. More specifically, it is possible in principle to perform arbitrary quantum computations by scattering wave packets through a graph, with interactions governed by the Hubbard model. 

The approach we have chosen to encode these computations is to simulate the Heisenberg Hamiltonian with time-varying control through the use of discrete scattering events which each have the same effect as a short period of Heisenberg interaction. As a simple special case of this procedure, our method can therefore obviously be used to simulate the Heisenberg Hamiltonian on a line or lattice.

In the case of strong coupling at half-filling, there is only reflection and no transmission in the scattering events, as would be the case for classical hard spheres. This provides an amusing echo of the Fredkin-Toffoli billiard ball computer that played an important role in the history of reversible computation~\cite{fredkin1982conservative}. Because the particles involved here are indistinguishable, however, the distinction between transmission and reflection is actually somewhat ambiguous.

The computational universality demonstration in this article is a proof of principle rather than a proposal for a concrete realization of quantum computation. Indeed, by prohibiting any time-varying control in the experiment, we have specifically eliminated one of the experimentalist's most powerful tools. As a consequence, while the overhead in performing quantum computation this way is polynomial, the degree of the polynomial is intimidating, scaling as the thirteenth power of the number of qubits and the fifth power of the number of logical gates. That being said, the estimates provided here for the overhead involved in performing the quantum computation are almost certainly far too conservative. The analysis ignores likely cancellations between many forms of errors, for example, and uses square wave packets for simplicity even though Gaussian wave packets exhibit less dispersion. Various tricks, such as incorporating the ability to perform gates corresponding to the two-dimensional representation of $S_3$ by permuting the three rails of the logical qubit, could also result in significant savings. It is even conceivable that better analysis and a more clever encoding of the computation could yield enormous savings: perhaps such an approach to quantum computation could even be made practical with the inclusion of some limited spatial inhomogeneity or time-varying control. 

\medskip

\acknowledgments
We thank John Joseph Carrasco, Andrew Childs, Steve Shenker and Brian Swingle. This research was supported by the Canadian Institute for Advanced Research and the Simons Foundation. This material is based upon work supported by the National Science Foundation Graduate Research Fellowship under Grant No. DGE-114747, and the Natural Sciences and Engineering Research Council of Canada Postgraduate Scholarship program.

\bibliography{references}
\end{document}